\documentstyle[12pt]{article}
\newtheorem{theorem}{Theorem}

\newtheorem{corollary}[theorem]{Corollary}
\newtheorem{definition}[theorem]{Definition}
\newtheorem{remark}[theorem]{Remark}
\newtheorem{example}[theorem]{Example}

\def\be{\begin{equation}}
\def\ee{\end{equation}}

\def\be{\begin{equation}}
\def\ee{\end{equation}}

\def\be{\begin{equation}}
\def\ee{\end{equation}}

\def\R{{\bf R}} 
 
\def\N{{\bf N}}
\def\T{{\bf T}}  
\def\C{{\bf C}}
\def\Re{{\rm Re}\,}

\def\H{\cal H}

\def\ep{\epsilon}

\def\PT{{\cal P}{\cal T}}
\def\P{{\cal P}}
\def\T{{\cal T}}

\def\l{{\lambda}}

\date{}
\begin{document}
\baselineskip=18pt
\title{Perturbation theory of $\PT$ symmetric Hamiltonians}

\author{
E. Caliceti$^1$, F. Cannata$^2$ and S. Graffi$^1$}
\date{}
\maketitle 
\hspace{-6mm}$^1$Dipartimento di Matematica, Universit\'a di Bologna, and INFN, Bologna,Italy\\
$^2$Dipartimento di Fisica and INFN, Via Irnerio 46,40126 Bologna,Italy
 
\begin{abstract}
 { \noindent
In the framework of perturbation theory the reality of the perturbed eigenvalues of a class of $\PT$symmetric Hamiltonians is proved using stability techniques. We apply this method to $\PT$symmetric unperturbed Hamiltonians perturbed by $\PT$symmetric additional interactions.}
\end{abstract}
\vskip 12pt\noindent 
\section{Introduction}
\setcounter{equation}{0}%
\setcounter{theorem}{0}%
Perturbation theory has played a very important role in the past (\cite{CGM}) in the study of non
Hermitian Hamiltonians with $\PT$ symmetry ({\cite{PT}). In \cite{CGS1, CGS2, Calogero} it was applied
starting from a selfadjoint Hamiltonian in order to investigate the perturbation of its spectrum by a
$\PT$ symmetric interaction. In the present paper we try to present more general results concerning
in general the perturbation of a non selfadjoint $\PT$
symmetric Hamiltonian:
$$ 
H \neq H^*\,,\quad H\PT = \PT H\,.
$$
This extension of perturbation theory will be shown to have non trivial aspects even when one
restricts oneself to unperturbed discrete simple levels with real energies. Our aim is to
provide a consistent and selfcontained framework where well established results and new
developments can be discussed coherently. The main original aspects that we present are:
\begin{enumerate}
\item
extension of perturbation theory for $\PT$ symmetric Hamiltonians in order to prove the
reality of the spectrum of a class of $\PT$ symmetric operators perturbed by a $\PT$ symmetric
interaction;
\item
the interaction does not need to be bounded relative to the unperturbed Hamiltonian;
\item
the results concerning the reality of the eigenvalues are achieved by using the stability
theory developed by  Hunziker  and Vock (HV-stability theory) in \cite{HV}.
\end{enumerate}
The paper is organized as follows. In Section 2 we give a general presentation of
perturbation theory for non selfadjoint $\PT$ symmetric operators and a review of the
stability theory  for eigenvalues. In Section 3 we give the technical results leaving the
proof to Section 4. Some open problems and further perspectives are outlined in the
Conclusions (Section 5).
\section{General formalism in perturbation theory and review of results}
\setcounter{equation}{0}%
\setcounter{theorem}{0}%
An operator is called $\PT$ symmetric if it  is invariant under the
combined action of a reflection operator $\P$ and the antilinear complex conjugation operator
$\T$. One basic issue is  to prove the reality of the spectrum. Results in this direction
have been obtained in \cite{Dorey, Shin} by means of ODE techniques, and they have been
recently extended in \cite {CGS1, CGS2, Calogero} (see also \cite{Caliceti} for a brief review) in the
framework of perturbation theory by perturbing selfadjoint Hamiltonians. The main goal of this paper is
to extend part of these results to the case of $\PT$ symmetric Hamiltonians obtained by perturbing a
$\PT$ symmetric operator, not necessarily selfadjoint. Thus, for the convenience of the reader, we first
recall the results of \cite {CGS1, CGS2, Calogero}, concerning $\PT$ symmetric operators in a Hilbert
space ($\H$,
$<u,v>$) of the form 
$$
H_g = H_0 + igW\,, \quad g\in\R\,,
$$
where $H_0$ is selfadjoint, with domain $D\subset\H$, and $W$ is a symmetric operator,
relatively bounded with respect to $H_0$. Moreover there exist a unitary  involution
$\P:\H\to\H$ and an antilinear involution $\T:\H\to\H$, both mapping $D$ to $D$, such that
\be
\label{PT}
H_0\P = \P H_0, \; \P W = -W\P,\; \T H_0 = H_0\T, \; \T W = W\T\,.
\ee
Then $H_g$ is $\PT$ symmetric, i.e. $\PT H_g = H_g\PT$, for $g\in\R$. An example is provided
by $\PT$ symmetric Schr\"o\-dinger operators in $L^2(\R^d)$, $d\geq 1$, where $H_0$ is the
selfadjoint realization of $-\Delta + V$, $\T$ is complex conjugation, i.e. $(\T\psi)(x) =
\overline{\psi(x)}$, and $\P$ is the parity operation defined by 
\be\label{P}
(\P\psi)(x) = \psi((-1)^{j_1}x_1,...,(-1)^{j_d}x_d)\,,\quad \psi\in L^2(\R^d)
\ee
where $j_k =0,1$, and $j_k =1$ for at least one $1\leq k\leq d$. Here $-\Delta$ denotes the
$d$-dimensional Laplace operator; $V$ and $W$ are real valued functions, $\P$-even and
$\P$-odd respectively: $\P V =V, \P W =-W$. The following theorem, proved in \cite{CGS1, CGS2},
provides a result in the case of bounded perturbation.  
\begin{theorem} \label{T2.1}
Let $H_0$ be a selfadjoint operator in $\H$ and $W$ a symmetric
operator in $\H$, satisfying the above assumption (\ref{PT}). Assume further
that
$H_0$ is bounded below, $W$ is bounded and that the spectrum of $H_0$ is
discrete. Let $\sigma(H_0) = \{E_j:j=0,1,\dots\}$ denote the incresing
sequence of distinct eigenvalues of $H_0$. Finally, let $\delta:=
\inf_{j\geq 0}\; [E_{j+1}-E_j]/2$ and assume that $\delta >0$. Then the following results
hold:
\begin{enumerate}
\item[(i)]
if for each degenerate eigenvalue of $H_0$ the corresponding eigenvectors have the same
$\P$-parity, i.e. they are either all $\P$-even or all $\P$-odd, then $\sigma(H_g)\subset\R$,
if $g\in\R$, $|g|<\delta/\|W\|$;
\item[(ii)]
if $H_0$ has an eigenvalue $E$ with multiplicity $2$ whose corresponding eigenvectors have
opposite $\P$-parity, i.e. one is $\P$-even and the other one is $\P$-odd, then $H_g$ has a
pair of non real complex conjugate eigenvalues near $E$ for $|g|$ small, $g\in\R$.
\end{enumerate}
\end{theorem}
The proof of the reality of the spectrum under suitable conditions has been extended to the
case of relatively bounded perturbation in \cite{Calogero} and is recalled in the following 
\begin{theorem} \label{T2.2}
Let 
$$
H_g = -\frac{d^2}{dx^2} + V(x) + ig W(x) \quad {\rm in}\; L^2(\R)\,,
$$
where $V(x)$ is a real-valued even polynomial of degree $2l$, with $\lim_{|x|\to
\infty}V(x) =+\infty$ and $W(x)$ is a real-valued odd polynomial of degree $2r-1$, with
$l>2r$. Then there exists $g_0>0$ such that $\sigma(H_g)\subset\R$ for $|g|<g_0$.
\end{theorem}
\begin{remark} \label{R2.3}
{\rm The above theorems enlarge the class of $\PT$symmetric Hamiltonians with real spectrum provided in \cite{Dorey, Shin}.}
\end{remark}
The question that we want to address in this paper is the folllowing. Let now $H_0$ be $\PT$
symmetric (not necessarily selfadjoint) with real spectrum and let $W$ be $\PT$
symmetric as well. We will provide conditions on $H_0$ and $W$ in order to guarantee that the
perturbed eigenvalues of $H(\ep):= H_0 + \ep W$, $\ep$ real, are real for $|\ep|$ small.
Therefore we shall not prove that the whole spectrum of $H(\ep)$ is real, but that at least
the eigenvalues generated by the unperturbed real ones stay real. First of all notice that if
$H_0$ has degenerate eigenvalues there may be problems even in the selfadjoint case, as
stated in Theorem \ref{T2.1}(ii). In general we have 
\be\label{M0}
 m_g(\lambda) \leq m_a(\lambda)
\ee
where
\be\label{M1}
m_g(\lambda) = {\rm dim}\{u : (H_0 -\l)u =0\}
\ee
is the geometric multiplicity of $\l\in\C$ and 
\be\label{M2}
m_a(\lambda) = {\rm dim}\{u: (H_0-\lambda)^n u=0\,, {\rm for\, some}\;n\in\N\}
\ee
is the algebraic multiplicity of $\l$. Notice that in the finite
dimensional case (i.e. $H_0$ and $W$ are matrices) $m_a(\lambda)$ is the
multiplicity of $\l$ as a root of the characteristic polynomial of $H_0$.
If $H_0$ is selfadjoint then $m_g(\lambda) = m_a(\lambda)$, for all
$\l\in\C$. However in the general (non selfadjoint) $\PT$symmetric case it
is not enough to assume that $m_g(\l)=1$ in order to guarantee that the
eigenvalues are simple, i.e. that there is no degeneracy, i.e. that
$m_a(\l)=1$. In fact there may be "exceptional points", caused by the
presence of Jordan blocks.
\begin{example}\label{E2.4}
{\rm Set
\be\label{Matrix}
H_0=\left(\begin{array}{ll} 
1 & i
\\
i & -1
\end{array}\right)
\,,\quad
\P=\left(\begin{array}{ll} 
1 & 0
\\
0 & -1
\end{array}\right)
\ee
and let $\T$ be complex conjugation. Then $H_0$ is $\PT$ symmetric. Its characteristic
polynomial ${\rm det}(H_0-\lambda I)= \lambda^2$ has just one root $\l_0 = 0$ with
$m_g(\lambda_0)=1$ and $m_a(\lambda_0)=2$. If we take 
$$ 
W=\left(\begin{array}{ll} 
0 & i
\\
i & 0
\end{array}\right)
$$
as a perturbation, then $H(\ep) = H_0 + \ep W$ has a pair of non-real complex conjugate
eigenvalues $\l(\ep) = \pm i\sqrt{\ep (\ep +2)}$ for $\ep >0$.}
\end{example}
To avoid such difficulties we analize here only the non-degenerate case, i.e. we assume that
\be\label{simple}
m_a(\lambda) = 1
\ee
for all eigenvalues $\l$ of $H_0$. The Hamiltonians $H_0$ that we perturb are
one-dimensional $\PT$symmetric Schr\"odinger operators with real simple spectrum, for example
the Hamiltonians $H_g$ provided by the above Theorems \ref{T2.1} and \ref{T2.2} for $|g|$
small.
\begin{remark}\label{R2.5}
{\rm Perturbation theory for matrices in the non-degenerate case provides a straightforward
result because: 
\begin{enumerate}
\item[(1)]
the stability of the unperturbed eigenvalues (see Definition \ref{D2.6} below) is guaranteed
by the boundedness of the perturbation $W$ (any matrix $W$ is a  bounded operator). This
implies that near any unperturbed eigenvalue $\l$ of $H_0$ there is one and only one 
eigenvalue
$\l(\ep)$ of
$H(\ep)$ for $|\ep|$ small and $\lim_{\ep\to 0}\l(\ep) = \l$. Thus $m_a(\lambda(\ep)) = 1$;
\item[(2)]
the eigenvalues of $\PT$symmetric operators come in pairs of complex conjugates, i.e. if 
$\l(\ep)$ is an eigenvalue of $H(\ep)$, then $\overline{\l(\ep)}$ is an eigenvalue of
$H(\ep)$ too.
\end{enumerate}
Therefore, combining (1) and (2), $\l(\ep)\in\R$.}
\end{remark}
We will provide new results on the reality of the perturbed eigenvalues ($\l(\ep)$) in the
case of unbounded (not necessarily relatively bounded) perturbation. As observed in Remark
\ref{R2.5}, a crucial issue is the stability of the unperturbed eigenvalues, i.e. the
problem is reduced to a stability result for the eigenvalues of $H_0$. Such result is immediate
if the perturbation $W$ is bounded relative to  $H_0$ (see e.g. \cite{Kato, RS}). Indeed, in
this case the  Rayleigh-Schr\"odinger perturbation expansion (RSPE) is convergent, for
$|\ep|$ small, to the (real) perturbed eigenvalues. However there can be stability even if
the RSPE does not exist. We provide here a brief reminder on stability (see \cite{Kato} and
\cite{HV}).
\begin{definition}
\label{D2.6}
A discrete eigenvalue $\l$ of $H_0$ is
stable with respect to (w.r.t.) the family 
$H(\ep) = H_0+\ep W$ if
\begin{enumerate}
\item[(i)]
for any small $r>0$,
$$
\Gamma _r = \{z:\mid z-\l\mid = r\}\subset\rho(H(\ep)) \,, \quad {\rm as}\; \ep\to 0
$$
where $\rho(H(\ep)):=\C-\sigma(H(\ep))$ is the resolvent set of $H(\ep)$;
\item[(ii)]
 $ \lim_{\ep\to 0} \parallel P(\ep) - P(0)\parallel = 0$
where
$$
P(\ep) = (2\pi i)^{-1} \oint _{\Gamma _r} (z - H(\ep))^{-1}dz
$$
is the spectral projection of $H(\ep)$ corresponding to the part of the
spectrum enclosed in $\Gamma _r$, and $H(0):=H_0$.
\end{enumerate}
\end{definition}
Since (ii) implies that 
\be\label{dim}
 {\rm dim} P(\ep) = {\rm dim}P(0) \;(=m_a(\l))
\ee
for $|\ep|$ small, $\l$ is the limit of a group of perturbed eigenvalues with the same total
algebraic multiplicity (see \cite{Kato}). If $W$ is bounded relative to $H_0$ there exist $a,
b>0$ such that
\be\label{relative}
\parallel Wu\parallel \leq b\parallel H_0u\parallel + a\parallel u\parallel \,,\quad\forall
u\in D(H_0).
\ee
Then
\be\label{NRC}
\parallel (z_0 - H(\ep))^{-1} - (z_0 - H_0)^{-1}\parallel \to 0\,,\quad {\rm as}\; \ep\to 0
\ee
for some $z_0\notin\sigma(H_0)$ and this implies (ii).
\begin{remark}
\label{R2.7}
{\rm Eigenvalues may be stable even if (\ref{NRC}) fails. As an example let $H(\ep) = p^2 +
x^2 + \ep x^4$, $\ep>0$, in} $\H$ $=L^2(\R)$ {\rm be the Hamiltonian of the even
anharmonic oscillator. Here $p^2=-d^2/dx^2$. Then the eigenvalues of the harmonic oscillator 
$H(0)$ are stable w.r.t. $H(\ep)$, $\ep\geq 0$, in spite of the fact that (\ref{NRC}) fails and the RSPE is divergent. In other words, the continuity of the eigenvalues at $\ep=0$ holds although analitycity fails. In
this case we only have
 strong resolvent convergence, i.e.:
\be\label{SRC}
\lim_{\ep\to 0}\, (z_0 - H(\ep))^{-1}u = (z_0 - H(0))^{-1}u\,,\quad \forall u\in\H
\ee
for some $z_0\notin\sigma(H(0))$. Then (\ref{SRC}) yields the strong convergence of the
projections $P(\ep)$, i.e.
\be\label{SPC}
\lim_{\ep\to 0} P(\ep)u = P(0)u\,,\quad \forall u\in\H 
\ee
which implies} 
\be\label{dimINEQ}
{\rm dim} P(\ep) \geq {\rm dim}P(0) \,,\quad {\rm as}\; \ep\to 0\,.
\ee
{\rm So $H(\ep)$ may have more eigenvalues than $H(0)$ in the circle $\Gamma_r$. This happens
for instance for the double-well operator
\be\label{DW}
H(\ep) = p^2 + x^2(1-\ep x)^2\,, \quad {\rm in}\; L^2(\R).
\ee
In fact near any eigenvalue of the harmonic oscillator $H(0)$ there are two eigenvalues of
$H(\ep)$ for $|\ep|$ small.}
\end{remark}
\section{Statement of the results}
\setcounter{equation}{0}%
\setcounter{theorem}{0}%
Let $H(\ep) = p^2 + V + \ep W, \ep \geq 0$, denote the closed operator in $L^2(\R)$ with
$C_0^{\infty}(\R)$ as a core, defined by
\be\label{def}
H(\ep)u = -u'' + Vu + \ep Wu\,,  \quad \forall u\in D(H(\ep))\,,
\ee
where $V=V_+ + iV_-$, $W=W_+ + iW_-$ and $V_+, V_-, W_+, W_-$ are real valued functions in
$L^{\infty}_{loc}(\R)$. Moreover $V_+, W_+$ are $\P$-even and bounded below, $V_-, W_-$ are
$\P$-odd and
\be\label{LIMIT}
 \lim_{\stackrel{|x|\to\infty}{\ep\to 0}}|V(x)+\ep W(x)| =+\infty.
 \ee
Here $(\P u)(x)=u(-x), \forall u\in
L^2(\R)$. Let us further assume that the spectrum of $H(\ep)$, denoted $\sigma(H(\ep))$, is discrete for $\ep\in[0,\ep_0]$, i.e.
it consists of a sequence of isolated eigenvalues with finite algebraic multiplicity:
\be\label{spectrum}
\sigma(H(\ep)) = \{ E_j(\ep): j=0,1,\dots\}.
\ee
For the unperturbed eigenvalues $(E_j(0))$ we will adopt the simplified notation
$E_j:=E_j(0), j=0,1,\dots$
\begin{theorem}
\label{T3.1}
Under the above assumptions the following statements hold:
\begin{enumerate}
\item[(1)]
each eigenvalue $E_j$ of $H(0)$ is stable w.r.t. the family $H(\ep), \ep>0$. In particular, if 
$E_j$ is simple, i.e. $m_a(E_j) =1$, and real there exists $\ep_j>0$ such that for $0<
\ep<\ep_j$, $H(\ep)$ has exactly one eigenvalue $E_j(\ep)$ close to $E_j$:
$$
\lim_{\ep\to 0} E_j(\ep) = E_j
$$
and $E_j(\ep)$ is real;
\item[(2)]
there are no "dying eigenvalues", i.e. if $E(\ep)\in\sigma(H(\ep))$ and $\lim_{\ep\to 0}
E(\ep) = E$, then $E$ is an eigenvalue of $H(0)$.
\end{enumerate}
\end{theorem}
In the following examples all the above conditions are satisfied. Moreover, the eigenvalues
of $H(0)$ are real and simple.
\begin{example}
\label{E3.2}
{\rm The unperturbed Hamiltonian $H(0)$ can be any of the operators $H_g$ satisfying the
above  Theorems \ref{T2.1} and \ref{T2.2}, for instance $H(0) = p^2+x^{2n}+ig\sin x$, $H(0)=
p^2+x^{2n}+igx/(x^2+1)$, $H(0) = p^2+x^{2l}+igx^{2q-1}$, with $|g|$ small, $n,l,q\in\N$ and
$l>2q$. Another example is provided by $H(0) = p^2+ix^{2k+1}$, $k\in\N$.
In all these cases the perturbation $W$ can be taken of the form $W=W_++iW_-$ where $W_+ =
\exp{(x^2)}$, or $W_+$ is an even polynomial function diverging positively at infinity, and
$W_-$ is an odd polynomial function bounded relative to $W_+$. If $H(0) = p^2+ix^{2k+1}$, $k\in\N$, the polynomial $W_-$ must diverge positively at $+\infty$ in order to guarantee that condition (\ref{LIMIT}) is satisfied.}
\end{example}
\begin{corollary}
\label{C3.3}
Assume that the eigenvalues $E_j, j\in\N$, of $H(0)$ are all simple and real. Then all the
perturbed eigenvalues $E_j(\ep), 0<\ep<\ep_j$, of $H(\ep)$ are real and simple. Moreover
(non-real) complex eigenvalues of $H(\ep)$ cannot accumulate at finite points but only at
infinity.
\end{corollary}
\begin{remark}
\label{R3.4} 
\begin{enumerate}
\item
{\rm The question of the reality of the whole spectrum of $H(\ep)$ is still open,
because there may be (possibly complex) eigenvalues of $H(\ep)$ diverging to infinity as
$\ep\to 0$}. 
\item
{\rm Hamiltonians satisfying Corollary \ref{C3.3} are provided in the above Example
\ref{E3.2}}.
\end{enumerate}
\end{remark}
If the perturbation $W$ is bounded relative to $H(0)$ then the RSPE near any $E_j$ converges.
We have therefore the following
\begin{corollary}
\label{C3.5}
If the unperturbed eigenvalues $E_j$ are simple and real and $W$ is bounded relative to
$H(0)$, then the RSPE near $E_j$ is real for all $j=0, 1,\dots$. More precisely, for $0<\ep
<\ep_j$
\be\label{RSPE}
E_j(\ep) = \sum _{n=0}^{\infty}a_n \ep ^n\;,\quad a_0=E_j
\ee
and $a_n\in\R, \forall n\in\N$.
\end{corollary}
\begin{remark}
\label{R3.6}
{\rm If $W$ is not relatively bounded w.r.t. $H(0)$, the RSPE although divergent can be Borel
summable to $E_j(\ep)$. In this case the coefficients of the RSPE are real. This happens for
instance for the Hamiltonians $H(0)$ and the perturbations $W$ of polynomial type described
in Example \ref{E3.2}}.
\end{remark}
\begin{remark}
\label{R3.7}
{\rm The stability theory developed in \cite{HV} allows one to prove a result similar to that
stated in Theorem \ref{T3.1} also in presence of essential spectrum provided that}
\be
\label{ess}
{\rm dist}(E_j,\sigma _{ess}(H(\ep)))\geq c>0\;, \quad {\rm as}\; \ep\to 0.
\ee
{\rm where $\sigma _{ess}(H(\ep))$ is the complement in $\sigma (H(\ep))$ of the discrete spectrum of $H(\ep)$.}
\end{remark}
\section{Proof of Theorem \ref{T3.1}}
\setcounter{equation}{0}%
\setcounter{theorem}{0}%
Although it is a straightforward application of the HV-stability theory developed in \cite {HV}, for a pedagogical purpose and for the convenience of the reader we
describe here the main steps. The only "continuity condition" required by the HV-theory and
guaranteed by the assumptions of the theorem is:
\be\label{continuity}
\lim_{\ep\to 0} H(\ep)u = H(0)u\,, \quad \forall u\in C_0^{\infty}(\R).
\ee
Condition (\ref{continuity}) implies the strong convergence of the resolvents (\ref{SRC})
which yields (\ref{dimINEQ}) and is not enough, as already remarked, to ensure stability.
One key ingredient is the numerical range of $H(\ep)$:
\be\label{numerical}
N(\ep):=\{<u, H(\ep)u>: u\in D(H(\ep)), \|u\|=1\}
\ee
which contains the eigenvalues of $H(\ep)$ and  in the present case is contained in the right
half-plane:
\be\label{inclusion}
\sigma (H(\ep))\subset N(\ep)\subset {\cal R}_+:=\{z:\Re z\geq 0\}.
\ee
Next we need to introduce the set of uniform boundedness of the  resolvents:
$$
{\cal D}:= \{z: (z-H(\ep))^{-1}\, {\rm exists\, and \,is \,uniformly\, bounded\, as}\,\ep\to
0\}.
$$
The sets ${\cal D}$ and $N(\ep)$ are closely related to each other (see e.g. \cite{Kato}),
since $\C -N(\ep)\subset {\cal D}$. In particular for $z\in{\cal R}_-:=\{z:\Re z<0\}$, we 
have
\be\label{stima}
\|(z-H(\ep))^{-1}\|\leq {\rm dist}(z,N(\ep))^{-1}\leq |Re z|^{-1}\,,\quad \forall \ep>0.
\ee
Thus $z\in{\cal D}$. The HV-theory allows us to prove that ${\cal D}$ is much wider than
${\cal R}_-$. Indeed we can prove that the following alternative holds:
\begin{enumerate}
\item[(1')]
if $E\in\sigma(H(0))$ then $E$ is stable w.r.t. $H(\ep)$, $\ep>0$;
\item[(2')]
if $E\notin\sigma(H(0))$ then $E\in{\cal D}$.
\end{enumerate}
Thus, ${\cal D}$ coincides with the complement of the spectrum of $H(0)$. Notice that $(2')$
implies statement $(2)$ of the theorem. Moreover a first step in the proof of $(1')$ consists
in showing that  the circle $\Gamma_r$ is contained in ${\cal D}$, and this
guarantees the possibility to construct the projection $P(\ep)$ for all $\ep$ sufficiently
small (see the above Definition \ref{D2.6}). Finally, a third key ingredient, on which the
proofs of both $(1')$ and $(2')$ are based, is represented by the so called "characteristic
sequences" of $(z-H(\ep))$ or "Weyl-type sequences", i.e. sequences $(\ep_n,u_n)$ such that:
\be\label{Weyl}
\ep_n\to 0, \, u_n\in D(H(\ep_n)),\, \|u_n\|=1,\, u_n
\stackrel{w}{\to}0,\, \|(z-H(\ep_n))u_n\|\to 0\,.
\ee
Now it is
easy to check that for a suitable constant $a>0$ we have 
\be\label{energy}
<u,p^2u> \leq a(\Re <u,H(\ep)u> + <u,u>)\,,
\ee
$\forall u\in D(H(\ep))$, $\forall \ep\in[0,\ep_0]$. By making use  of (\ref{energy}) one can
prove that a sequence of type (\ref{Weyl}) generates another characteristic sequence
$(\ep_n,v_n)$ which is "supported at infinity", i.e. such that
\be\label{infinity}
v_n(x)=0\,, \quad {\rm for}\; |x|\leq n\,,\;\forall n\in\N.
\ee
Since $\lim_{|x|\to \infty} |V(x)| = +\infty$ by the above assumption (\ref{LIMIT}), we first assume that $\lim_{|x|\to \infty} V_+(x) =
+\infty$. Now the idea is to prove $(2')$, i.e. that $\sigma(H(0))\cup{\cal D}=\C$, by
contradiction. More precisely, one can show that if $z\notin\sigma(H(0))\cup{\cal D}$, then a
characteristic sequence $(\ep_n,v_n)$ of $(z-H(\ep))$ exists and it satisfies both
(\ref{Weyl}) and (\ref{infinity}). But this contradicts the fact that for all $z\in\C$
\be\label{distance}
\lim _{\stackrel{n \to\infty}{\ep\to 0}}d_n(z,\ep) = +\infty
\ee
where
\be\label{distanceBIS}
d_n(z,\ep) = \inf\{\|(z-H(\ep)): u\in D(H(\ep))\,,\|u\|=1\,,u(x)=0\,{\rm for}\; |x|\leq n\}\,.
\ee
Indeed we have
\be\label{distINEQ}
d_n(z,\ep)\geq{\rm dist}(z,N_n(\ep))
\ee
where
\be\label{rangeINFTY}
N_n(\ep)=\{<u, H(\ep)u>: u\in D(H(\ep)), \|u\|=1\,,u(x)=0\,{\rm for}\; |x|\leq n \}
\ee
is the so called "numerical range at infinity". Now, for all $u\in D(H(\ep))$ such that
$\|u\|=1$ and $u(x)=0$  for $|x|\leq n$, we have
\begin{eqnarray}\label{estimate}
|z- <u, H(\ep)u>|\;\geq\;Ê|<u, H(\ep)u>|-|z|  
\nonumber\\
\geq |\Re <u, H(\ep)u>|-|z|\;\geq \;<u,V_+u> - |z| -c
\end{eqnarray}
as $\ep\to 0$, for some constant $c>0$. Thus, (\ref{distance}) follows from (\ref{distINEQ})
and the fact that $\lim_{|x|\to \infty} V_+(x) = +\infty$ by assumption. Hence $(2')$ is
proved. Now let $ E\in\sigma(H(0))$. In order to prove that $E$ is stable w.r.t. $H(\ep)$,
$\ep>0$, we recall that it is enough to prove (\ref{dim}), i.e. in view of (\ref{dimINEQ}),
that 
\be\label{INEQ2}
{\rm dim} P(\ep) \leq {\rm dim}P(0) \,,\quad {\rm as}\; \ep\to 0\,.
\ee
Again the proof is by contradiction. In fact if we assume that ${\rm dim} P(\ep_n) > {\rm
dim}P(0)$, for $\ep_n\to 0$, a characteristic sequence of $(E-H(\ep))$ which 
satisfies (\ref{infinity}) can be found, and exactly as before this contradicts
(\ref{distance}). Thus the theorem is proved under the assumption that $\lim_{|x|\to \infty}
V_+(x) = +\infty$. If this assumption is not satisfied then it follows from (\ref{LIMIT}) that 
 $\lim_{\stackrel{|x|\to\infty}{\ep\to 0}}|V_-(x)+\ep W_-(x)| =+\infty$. Without loss of generality,
since $V_-$ and $W_-$ are $\P$-odd, we may assume that $\lim_{\stackrel{x\to +\infty}{\ep\to
0}}(V_-(x)+\ep W_-(x)) =+\infty$ and $\lim_{\stackrel{x\to -\infty}{\ep\to 0}}(V_-(x)+\ep W_-(x))
=-\infty$. This immediately implies that there exists $c\geq 0$ such that
$$
\lim _{\stackrel{n \to\infty}{\ep\to 0}}d_n(z,\ep)= 0\,, \quad \forall z\in{\cal R}_+\,,\, \Re z\geq c
$$ 
and therefore (\ref{distance}) cannot be used to generate a
contradiction with (\ref{Weyl}) and (\ref{infinity}). Nevertheless the problem is overcome as follows 
(see also \cite{CM} where an analogous problem was handled in a similar fashion). 
First of all one can prove that any characteristic sequence $(\ep_n,u_n)$ generates  another characteristic 
sequence  $(\ep_n,v_n)$ which is supported either at $+\infty$ or at $-\infty$, i.e. such that either 
\be\label{+infinity}
v_n(x)=0\,, \quad {\rm for}\; x\leq n\,,\;\forall n\in\N
\ee
or
\be\label{-infinity}
v_n(x)=0\,, \quad {\rm for}\; x\geq -n\,,\;\forall n\in\N\,.
\ee
Next we introduce the numerical range at
$+\infty$, $N_n^+(\ep)$, and the numerical range at $-\infty$, $N_n^-(\ep)$, defined by
(\ref{rangeINFTY}) where the condition $u(x)=0$ for $|x|\leq n$ is replaced by $u(x)=0$ for
$x\leq n$, and $u(x)=0$ for $x\geq -n$ respectively. More precisely:
$$
N_n^+(\ep)=\{<u, H(\ep)u>: u\in D(H(\ep)), \|u\|=1\,,u(x)=0\,{\rm for}\; x\leq n \}
$$
and
$$
N_n^-(\ep)=\{<u, H(\ep)u>: u\in D(H(\ep)), \|u\|=1\,,u(x)=0\,{\rm for}\; x\geq -n \}\,.
$$
Then for all $z\in\C$ we have
\be\label{distpm}
\lim _{\stackrel{n \to\infty}{\ep\to 0}}d_n^{\pm}(z,\ep) = +\infty
\ee
where
\be\label{dist+} 
d_n^+(z,\ep) = \inf\{\|(z-H(\ep)): u\in D(H(\ep))\,,\|u\|=1\,,u(x)=0\,{\rm for}\; x\leq
n\}\,.
\ee
and similarly for $d_n^-(z,\ep)$. Then the existence of a characteristic sequence satisfying either (\ref{+infinity}) or (\ref{-infinity}) contradicts (\ref{distpm}) and this concludes the proof of the theorem.
\section{Conclusions}
\setcounter{equation}{0}%
\setcounter{theorem}{0}%
The reality of the spectrum  of a $\PT$symmetric Hamiltonian has been  proved for a
class of polynomial Hamiltonians and for solvable classes of potentials generated by solvable
selfadjoint problems by a complex coordinate shift. By a suitable use of the  perturbation
theory we have summarized in this paper  that it is  conceivable  that the class of
$\PT$symmetric Hamiltonians with real spectrum can be considerably enlarged in so far as one
starts from a  $\PT$symmetric Hamiltonian with real spectrum and adds a $\PT$ symmetric
perturbation. This analysis does not  pretend to be  exhaustive in the sense  that the
perturbed eigenvalues of $H(\ep)$ are real but there is no guarantee that the whole
spectrum is real. Therefore the connection with pseudo-Hermiticity deserves further
investigation. Moreover the pseudo-Hermiticity condition
$H=\eta H^*\eta^{-1}$ (see e.g. \cite{MoJo, Figueira}) of the unperturbed Hamiltonian could generate a perturbative
approach to pseudo-Hermiticity for larger classes of Hamiltonians and of the operator  $\eta$. Another
open question is a detailed analysis of the degenerate case, i.e. the case when the unperturbed
eigenvalues are diabolic points (\cite{Kirillov}) or exceptional points (\cite{Seyranian, Weigert}); the
latter are typical of non diagonalizable Hamiltonians.
\section*{Acknowledgements}
We wish to thank D. Heiss, H. Geyer and M. Znojil for the warm hospitality at the workshop and for providing the opportunity to publish the contributions. We also thank P. Dorey for useful discussions and suggestions.
\vskip 1.5cm\noindent 


\begin{thebibliography}{DD} 

\bibitem[1]{CGM} 
Caliceti E, Graffi S and  Maioli M 1980 
{\it Commun. Math. Phys.} {\bf 75} 51  

\bibitem[2]{PT} 
Bender C M  and Boettcher S 1998 {\it Phys. Rev. Lett.} {\bf 80} 5243;
Cannata F, Junker G and Trost J 1998 {\it Phys. Lett.} A {\bf 246 }  219;
Andrianov A A, Cannata F, Dedonder J P and  Ioffe M  V 1999 {\it Int. J. Mod. Phys.}
A  {\bf 14} 2675;
Bender C M, Boettcher S and Meisinger P 1999 {\it J. Math. Phys.}  {\bf 40}
2201;
Bender C M, Milton K.A. and Savage V M 2000 {\it Phys. Rev.} D {\bf 62}  085001;
Bender C M, Brody D C and Jones H F 2004
{\it  Phys. Rev. Lett.} {\bf 93} 251601; Znojil M 1999 {\it Phys. Lett.} A {\bf 259 }  220

\bibitem[3]{CGS1}  
Caliceti E, Graffi S and Sj\"ostrand J 2005 {\it J. Phys. A:
Math. Gen.} {\bf 38} 185 

\bibitem[4]{CGS2} 
Caliceti E, Graffi S and Sj\"ostrand J: in preparation

\bibitem[5]{Calogero} 
Caliceti E and Graffi S 2005 {\it J. Nonlinear Math. Phys.} {\bf
12}, Supplement 1, 138

\bibitem[6]{HV} 
Hunziker W and Vock E 1982 
{\it Commun. Mathy. Phys.} {\bf 83} 281

\bibitem[7]{Dorey} 
Dorey P, Dunning C and Tateo R 2001 
{\it J. Phys. A: Matyh. Gen.} {\bf 34} L391

\bibitem[8]{Shin} 
Shin K C 2002 
{\it Commun. Math. Phys.} {\bf 229} 543

\bibitem[9]{Caliceti} 
Caliceti E 2004 {\it Czech. J. Phys.} {\bf 54}, n 10, 1065

\bibitem[10]{Kato}
Kato T 1976 {\it Perturbation Theory for Linear Operators},
2nd Edition (Berlin: Springer)

\bibitem[11]{RS} 
Reed M and Simon B 1978 {\it Methods of Modern Mathematical Physics} vol IV (New York:
Academic)

\bibitem[12]{CM}
Caliceti E and Maioli M 1983 {\it Ann. Inst. H. Poincar\'e} A {\bf 38} 175 

\bibitem[13]{MoJo} 
Mostafazadeh A 2002 {\it J. Math. Phys.} {\bf 43} 205; Mostafazadeh A 2003 {\it J. Phys. A: Math. Gen.}
{\bf 36} 7081; Jones H. and Mateo J. 2005 {\it Czech J. Phys.} {\bf 55}, n. 9, 1117; Mostafazadeh A
2005 {\it Czech J. Phys.} {\bf 55}, n. 9, 1157  

\bibitem[14]{Figueira} 
Figueira de Morisson Faria C and Fring A {\it Preprint} quant-ph 0604014; 
Bagchi B, Banerjee A and Quesne C {\it Preprint} quant-ph 0606012

\bibitem[15]{Kirillov}
Kirillov O and G\"unther U 2006 {\it Preprint} math-ph 0602013

\bibitem[16]{Seyranian}
Seyranian P A, Kirillov O N and Mailybaev A A 2005 {J. Phys. A: Math. Gen.} {\bf 38} 1723

\bibitem[17]{Weigert}
Weigert S 2005 {\it Czech J. Phys.} {\bf 55}, n. 9, 1183

\end{thebibliography}
\end{document}